\pdfoutput=1
\documentclass[10pt,conference,twocolumn,final]{IEEEtran}
\usepackage{amsmath,amsfonts}
\usepackage{booktabs}
\usepackage{graphicx}
\usepackage{bmpsize}
\usepackage{ifpdf}
\usepackage{tikz}
\usepackage{textcomp}
\usepackage{hyperref}
\ifCLASSOPTIONcompsoc
    \usepackage[caption=false,font=normalsize,labelfont=sf,textfont=sf]{subfig}
\else
    \usepackage[caption=false,font=footnotesize]{subfig}
\fi
\usepackage{url}

\newcommand{\sref}[1]{Section~\ref{#1}}

\graphicspath{
  {images/}
}
\ifpdf
\DeclareGraphicsExtensions{.pdf,.png}
\else
\DeclareGraphicsExtensions{.eps}
\fi

\IEEEoverridecommandlockouts

\newcommand\copyrighttext{%
	\footnotesize \textcopyright 2019 IEEE. Personal use of this material is permitted.
	Permission from IEEE must be obtained for all other uses, in any current or future
	media, including reprinting/republishing this material for advertising or promotional
	purposes, creating new collective works, for resale or redistribution to servers or
	lists, or reuse of any copyrighted component of this work in other works.
	DOI: \href{https://doi.org/10.1109/VTCSpring.2019.8746551}{10.1109/VTCSpring.2019.8746551}}
\newcommand\copyrightnotice{%
	\begin{tikzpicture}[remember picture,overlay]
	\node[anchor=north,xshift=0pt,yshift=-10pt] at (current page.north) {\fbox{\parbox{\dimexpr\textwidth-\fboxsep-\fboxrule\relax}{\copyrighttext}}};
	\end{tikzpicture}%
}

\begin{document}
\bstctlcite{ShortCTL:BSTcontrol}
\title{3.5 GHz Coverage Assessment with a 5G Testbed}

\newdimen\origiwspc%
\origiwspc=\fontdimen2\font%
\fontdimen2\font=0.7\origiwspc%

\author{
\IEEEauthorblockN{Adrian Schumacher\textsuperscript{*}\textsuperscript{\#},
    Ruben Merz\textsuperscript{*} and
    Andreas Burg\textsuperscript{\#}}
\IEEEauthorblockA{\textsuperscript{*}Enterprise Architecture \& Innovation, Swisscom Ltd., CH-3050 Bern, Switzerland\\
    \{adrian.schumacher,ruben.merz\}@swisscom.com}
\IEEEauthorblockA{\textsuperscript{\#}Telecommunications Circuits Laboratory, EPFL, CH-1015 Lausanne, Switzerland\\
    andreas.burg@epfl.ch\vspace{-0.5\baselineskip}}
}
\maketitle

\copyrightnotice

\begin{abstract}
  Today, cellular networks have saturated frequencies below 3\,GHz. Because of increasing capacity requirements, 5th generation (5G) mobile networks target the 3.5\,GHz band (3.4 to 3.8\,GHz).
Despite its expected wide usage, there is little empirical path loss data and mobile radio network planning experience for the 3.5\,GHz band available.
This paper presents the results of rural, suburban, and urban measurement campaigns using a pre-standard 5G prototype testbed operating at 3.5\,GHz, with outdoor as well as outdoor-to-indoor scenarios.
Based on the measurement results, path loss models are evaluated, which are essential for network planning.

\end{abstract}
\section{Introduction}
\label{sec:introduction}
The global demand for higher-capacity mobile Internet services drives the continuous evolution of mobile (cellular) technologies.
Up until the 4th generation mobile network (4G, also commonly known as Long Term Evolution (LTE)), only frequency bands up to around 2.7\,GHz have been used (with a few exceptions).
To carry the required increase in data capacity, more bandwidth in higher frequency spectrum is required.
CEPT/ECC identified\footnote{\url{https://www.cept.org/ecc/topics/spectrum-for-wireless-broadband-5g}}
the 3400--3800\,MHz frequency band to be harmonized in Europe for 5G usage, along with the 24.25--27.5\,GHz band. National regulators already have made, or are in the process of making the 3.4--3.8\,GHz (referred to as 3.5\,GHz) band available for mobile network operators. For instance, in Switzerland, an auction for this band was conducted during the first quarter of 2019.
With 5G, we refer to the Release~15 New Radio (NR) standard by the 3rd Generation Partnership Project (3GPP). 5G NR is frequency band agnostic and could be deployed on legacy bands (below 3\,GHz), but also in the milimeter-wave (mmWave) spectrum. As a compromise to still have fairly good propagation properties, but also allow for wider carrier bandwidths, the 3.5\,GHz band is of particular interest among the bands below 6\,GHz (sub-6\,GHz).
Propagation channels for mobile use in the 3.5\,GHz band and above, have mainly been studied in the context of WiMAX (Worldwide interoperability of Microwave Access) and with less carrier bandwidth compared to what 5G will be able to use (up to 200\,MHz carrier bandwidth for sub-6\,GHz). 
It also needs to be considered that an increase in bandwidth requires the same factor for a power increase, to maintain the same coverage area.

For radio network planning, propagation models are used, which can be categorized into deterministic, stochastic, empirical, and standardized models \cite[Chapter~4]{clerckx_mimo_2013}. While deterministic models (e.g., ray tracing) may yield accurate results, they require a high degree of detailed description and calibration of the environment and are computationally expensive. Stochastic models (e.g., COST~207) employ random variables and do not require much information about the environment, but also cannot provide high-accuracy path loss predictions. Empirical (e.g., Hata-Okumura) and standardized models are based on measurements and have been widely adopted to predict a mean path loss as a function of distance, frequency, and some further parameters specific to the environment or infrastructure. They require also only a few parameters and can provide an acceptable prediction accuracy, better than the very general and simplified path loss model (\ref{eq:simpl_pl_model}). For these reasons, radio network planners in the industry often use empirical path loss models for network planning simulations.\looseness=-1

\subsection{Overview of Empirical Path Loss Models} \label{sec:plmodels}

Because most mobile communications applications use frequencies up to around 2\,GHz, the empirical path loss models were optimized to cover only these frequencies. Examples are (see also Table~\ref{table:pl_model_list}):
Hata-Okumura \cite{hata_empirical_1980},
COST~231 Hata \cite{european_commission_cost_1999},
COST~231 Walfish-Ikegami (COST~231~WI) \cite{european_commission_cost_1999}, and
Erceg \cite{erceg_empirically_1999}.
Path loss models valid for the 3.5\,GHz band are the \emph{Stanford University Interim (SUI)} IEEE 802.16 model \cite{ieee_channel_2001} which is an extension to the Erceg model, the \emph{ECC-33} model \cite{ecc_analysis_2003} which extrapolates the Hata-Okumura model, the \emph{WINNER~II} models \cite{pekka_kyosti_winner_2008}, the \emph{ITU-R P.1411-9} models \cite{itur_p14119_2017}, as well as the \emph{3GPP} models described in \cite{zz3gpp.38.901}.

When designing wireless networks, it is obvious that underestimating the path loss can lead to coverage holes. However, also overestimating the path loss is undesirable because this leads to severe inter-cell interference issues. For an accurate prediction, existing models need to be evaluated for the environments in which they shall be used and adjusted if necessary. Additionally, penetration of signals from outdoor to indoor is depending heavily on the building materials and is varying a lot, even within the same cell coverage area. While wooden and older houses -- Rec. ITU-R P.2109-0 \cite{itur_p2109_2017} refers to `traditional buildings' -- tend to have a small penetration loss, modern `thermally-efficient' buildings with infrared light reflecting low emissivity (low-e) coated glass windows impose high losses.

\begin{table}[ht]
	\centering
	\caption{Selected Empirical Path Loss Models}
	\label{table:pl_model_list}
	\vspace{-0.8em}
	\begin{tabular}{@{}lcc@{}}
		\toprule
		Name & Frequency Range & Distance Range \\
		\midrule
		Hata-Okumura & $150-1500\,\mathrm{MHz}$ & $1-20\,\mathrm{km}$ \\
		COST 231 Hata & $1500-2000\,\mathrm{MHz}$ & $1-20\,\mathrm{km}$ \\
		COST 231 Walfish-Ikegami & $800-2000\,\mathrm{MHz}$ & $0.02-5\,\mathrm{km}$ \\
		Erceg & $\approx 2000\,\mathrm{MHz}$ & $0.1-8\,\mathrm{km}$ \\
		\hline
		SUI IEEE 802.16 & $1-4\,\mathrm{GHz}$ & $0.1-8\,\mathrm{km}$ \\
		ECC-33 & $3.4-3.8\,\mathrm{GHz}$ & $1-10\,\mathrm{km}$ \\
		WINNER II & $2-6\,\mathrm{GHz}$ & $0.05-5\,\mathrm{km}$ \\
		ITU-R P.1411-9 & $0.3-100\,\mathrm{GHz}$ & $0.055-1.2\,\mathrm{km}$ \\
		3GPP & $0.5-100\,\mathrm{GHz}$ & $0.01-5^{\mathrm{*}}\,\mathrm{km}$ \\
		\bottomrule
		$^{\mathrm{*}}$\footnotesize{10 km for RMa LOS} & & \\
	\end{tabular}
	\vspace{-1em}
\end{table}

In general the path loss (in dB) can be expressed according to the simplified model in \cite[eq.~(1.12)]{clerckx_mimo_2013}: 
\begin{equation}\label{eq:simpl_pl_model}
P_L(d) = A_0 + 10 \gamma \log_{10} \left( d/d_0 \right) + \chi_\sigma \quad d > d_0\mathrm{,}
\end{equation}
where $A_0$ represents the deterministic path loss component at the reference distance $d_0$ in meters. The path loss exponent is denoted $\gamma$ and $d$ is the distance in meters between base station (BS) and user equipment (UE).
Finally, $\chi_\sigma$ is the stochastic shadow fading component in dB with a zero-mean log-normal distribution and standard deviation $\sigma$. The above mentioned empirical models can also be expressed according to (\ref{eq:simpl_pl_model}) with additional terms depending on, e.g., the operating frequency, BS and/or UE antenna height, etc. \cite[eqs.~(3)--(12)]{imperatore_path_2007}.

\subsection{Related Work} \label{sec:related}

Before 5G, there was already an interest in the 3.5\,GHz band in the early 2000's with the radio access technology WiMAX for fixed wireless access. Therefore, several measurement results are available for the frequency range 3.4--3.8\,GHz, e.g., \cite{imperatore_path_2007, abhayawardhana_comparison_2005, walden_urban_2005, joseph_path_2006, kun_path_2008}. The corresponding measurement campaigns either used a WiMAX system with a signal bandwidth of 3.5\,MHz, or a continuous wave (CW) signal. Therefore, no conclusions can be drawn confidently for a wide bandwidth such as 100\,MHz that can be used for 5G. The BS antenna height varied between 15--36\,m, while the UE antenna height varied between 2--10\,m above ground. Some comparisons have been done with the models Erceg, COST~231~WI, COST~231 Hata, SUI, and ECC-33, but none with 3GPP's path loss models. All publications calculated a path loss exponent and standard deviation according to the simplified path loss model from (\ref{eq:simpl_pl_model}), which are summarized in Table~\ref{table:pl_result_comparison} (only the lowest UE height is considered in the table). The variations in these parameters (e.g., $\gamma$ for a suburban terrain ranges from 2.13 to 4.9 and for urban terrain from 2.3 to 4.3) indicate that there are many influences on the propagation channel that may depend on the geographic region, construction material, and vegetation which in parts has also been confirmed in \cite{riback_carrier_2006}. While \cite{abhayawardhana_comparison_2005} suggests the ECC-33 model for urban and the SUI (terrain B) for suburban environments, \cite{joseph_path_2006} found that the Erceg (terrain C) fits best, but underpredicts the measured path loss. For rural environments, \cite{imperatore_path_2007} and \cite{abhayawardhana_comparison_2005} show that the best fitting models SUI (terrain B \& C) and COST~231 Hata overestimate the measured path loss. Because the path loss model with the lowest prediction error varies, it is difficult to set on a specific model for network planning purposes.\looseness=-1

Regarding outdoor-to-indoor propagation, measurements were presented in \cite{rodriguez_path_2013}, along with few outdoor measurements.
A difference of only 10\,dB more attenuation was found for a modern building compared to an old building. According to the authors, this stems from different wall thicknesses and building material. Therefore we conclude that the modern building was not equipped with low-e coated windows.

\begin{table}
	\centering
	\caption{Comparison of Path Loss Parameters for (\ref{eq:simpl_pl_model}) at 3.5\,GHz}
	\label{table:pl_result_comparison}
	\vspace{-0.8em}
	\setlength\tabcolsep{4.5pt} 
	\begin{tabular}{@{}llllcc@{}}
		\toprule
		Terrain & Location & UE Height & Distance & $\gamma$ & $\sigma$ \\
		&  & [m] & [m] &  & [dB] \\
		\midrule
		rural 	& Cambridge, UK \cite{abhayawardhana_comparison_2005}	& 6 	& 250-2000 	& 2.7 	& 10\\
		rural 	& Piemonte, Italy \cite{imperatore_path_2007}			& 2 	& 1000-10000& 2.5 	& 8.9\\
		suburban 	& Cambridge, UK \cite{abhayawardhana_comparison_2005}& 6 	& 250-2000 	& 2.13 	& 11.1\\
		suburban 	& Ghent, Belgium \cite{joseph_path_2006}			& 2.5 	& 30-1500 	& 4.9 	& 7.7\\
		suburban 	& Shanghai, China \cite{kun_path_2008} 				& 3 	& 300-1800 	& 3.6 	& 9.5\\
		urban  	& Cambridge, UK \cite{abhayawardhana_comparison_2005}	& 6 	& 250-2000 	& 2.3	& 11.7\\
		urban 	& United Kingdom \cite{walden_urban_2005}				& 2.5 	& 100-2000 	& 4.3 	& 7.5\\
		\bottomrule
	\end{tabular}
	\vspace{-0.5em}
\end{table}

\subsection{Contribution and Outline}
The related works listed above show that the path loss characteristics (path loss exponent, standard deviation) vary a lot depending on the environment. Because it is difficult to decide for a specific path loss model and do the network planning accordingly, we conducted extensive measurements in Switzerland in a rural, suburban, and urban environment.
The measurement setup and environments are described in \sref{sec:setup}. Contrary to most prior works, a beamforming BS antenna was used and parallel measurements on a live network in legacy frequency bands were conducted for comparison
and validation, and more realistic UE antenna heights for mobile cellular applications were used.
The obtained results are compared against the 3GPP, WINNER~II, and SUI path loss models, and described in \sref{sec:measresults} (ITU-R P.1411-9 was excluded due to the specificity to environments).
We find that most models overestimate the path loss, and for every scenario, a different model predicts the path loss with the least error.
Finally, the paper is concluded in \sref{sec:conclusions}.

\section{Measurement Setup and Environment}
\label{sec:setup}

\subsection{Measurement Method}
The propagation path loss in the downlink is defined as the radio frequency (RF) attenuation $P_L$ in dB of a transmitted signal when it arrives at the receiver:
\begin{equation}\label{eq:pl_meas_based}
P_L = P_T + G_T(\phi,\theta) + G_R - P_R \quad [\mathrm{dB}]\mathrm{,}
\end{equation}
where $P_T$ is the BS transmitted power in dBm and $G_T(\phi,\theta)$ the transmitter antenna gain in dB as a function of azimuth angle $\phi$ and elevation angle $\theta$. Because the UE antenna is omnidirectional, we can simplify the receiver antenna gain as a constant $G_R$ in dB. Finally, $P_R$ is the local mean received power in dBm.

For the analysis, all measurement parameters were geographically binned in a two-dimensional square 5\,m grid, thus removing fast fading effects according to the Lee sampling criterion \cite{lee_estimation_1974}, and to prevent a bias from temporal influences due to, e.g., stops at red traffic lights. For each bin, the median value was computed.
For the log-distance path loss plots, also the median value was calculated for each 5\,m distance bin.

\subsection{5G Testbed}
\begin{figure}[t]
	\centering
	\vspace{-1em}
	\subfloat[5G Testbed\label{fig:testbed}\vspace{-1em}]{%
		\hspace*{\fill}\includegraphics[width=0.47\linewidth]{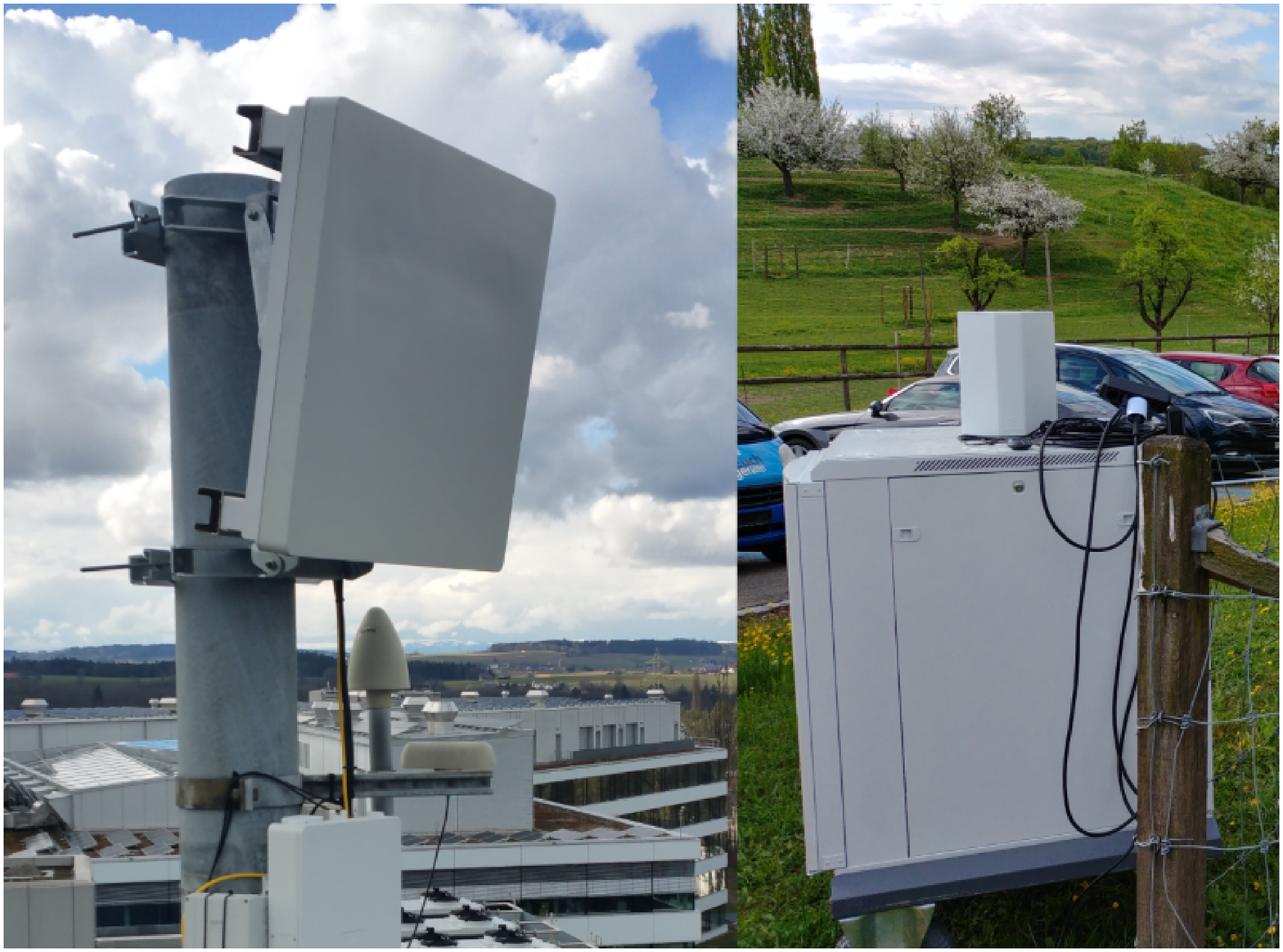}}
	\hfill
	\subfloat[Rural Deployment\label{fig:map_meikirch}\vspace{-1em}]{%
		\includegraphics[width=0.49\linewidth]{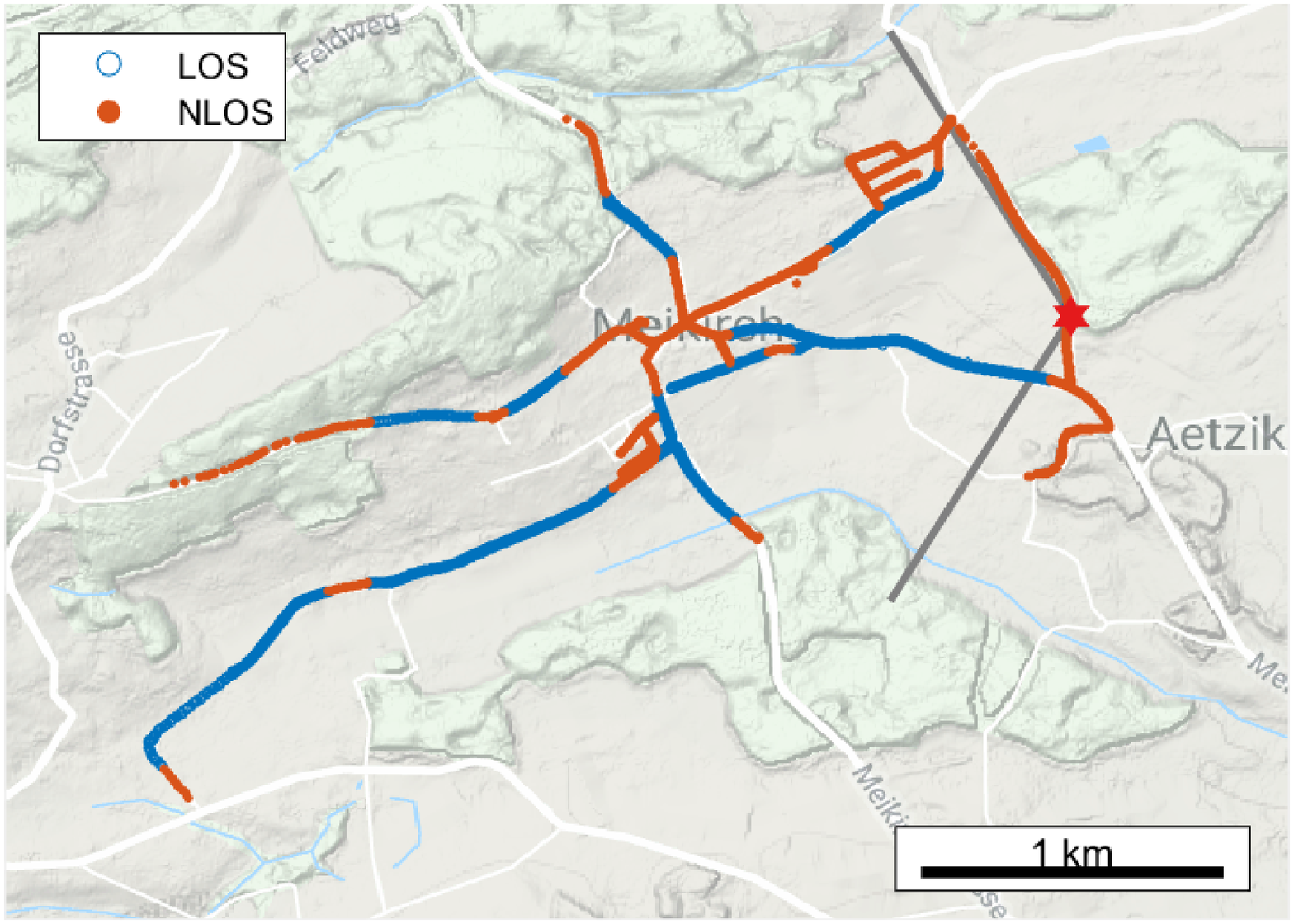}}
	\\
	\subfloat[Suburban Deployment\label{fig:map_ittigen}]{%
		\includegraphics[width=0.49\linewidth]{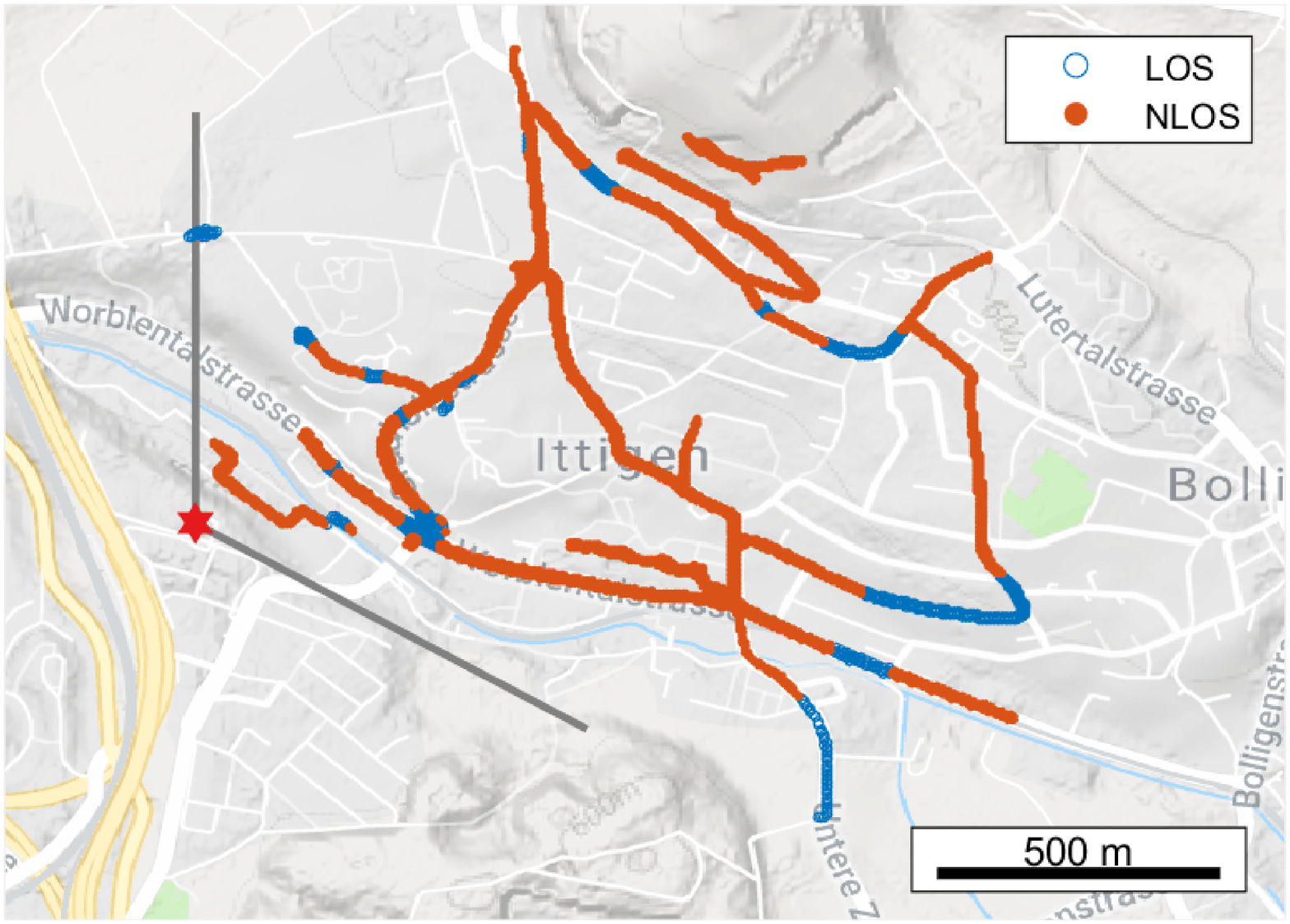}}
	\hfill
	\subfloat[Urban Deployment\label{fig:map_zurich}]{%
		\includegraphics[width=0.49\linewidth]{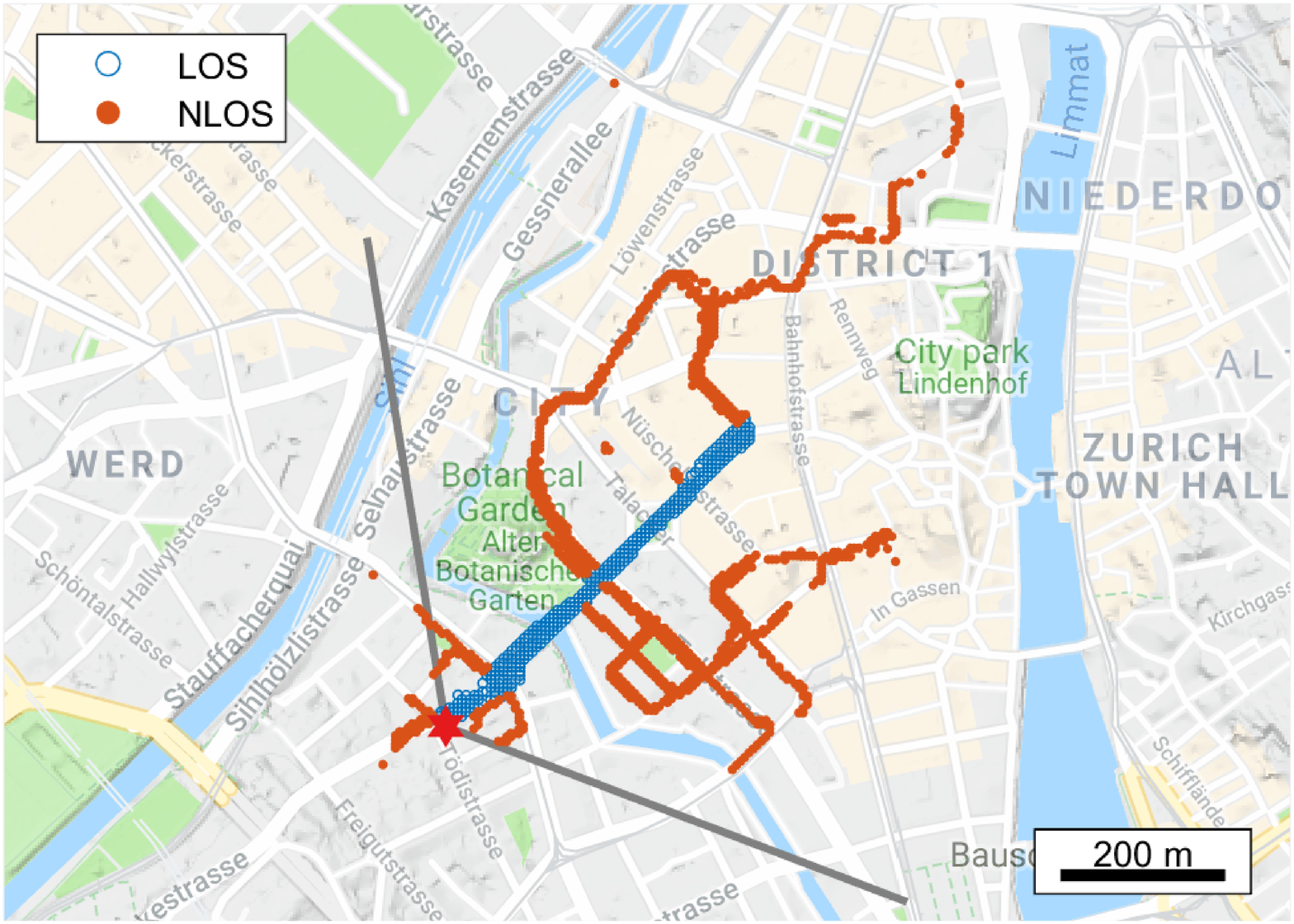}}
	\caption{(a) Shows the 5G testbed AAS on a mast and the UE in the field, (b) the map of the rural deployment with a red star indicating the AAS location with the 120$^\circ$ sector, and the LOS/NLOS classification. Similarly, (c) and (d) show the map of the suburban and urban deployments, respectively.
		\label{fig:testbedAndMaps}} 
\end{figure}

For conducting measurements as close to 5G as possible, we employed a 5G testbed (similar to \cite{halvarsson_5g_2018}). It consists of a BS unit, an active antenna system (AAS) connected via fiber to the BS, and in our case two UEs (see \figurename~\ref{fig:testbed} for a picture of the AAS and one UE).
The center frequency for which test-licenses were available, was 3.55\,GHz. The time division duplex (TDD) and orthogonal frequency division multiplexing (OFDM) signal occupies a bandwidth of 80\,MHz.
Further details can be found in \cite[Tab.~1]{halvarsson_5g_2018} and the references therein. Contrary to the beam grid described in \cite{halvarsson_5g_2018}, we were using a configuration with the 48 different dual-polarized beams arranged in a grid of three rows with 16 beams each (visible in \figurename~\ref{fig:antenna_patterns}). The antenna gain and resulting coverage remains approximately the same as with the configuration described in \cite{halvarsson_5g_2018} with 27\,dBi, and 120$^\circ$ in azimuth and 30$^\circ$ in elevation, respectively.
The two identical UEs have eight antenna ports connected to an eight element omnidirectional cross-polarized antenna, each element with a gain of 6\,dBi.\looseness=-1

As in \cite{halvarsson_5g_2018}, the Mobility Reference Signal Received Power (MRSRP) for each of the 48 beams has been logged and used for the analysis. The Mobility Reference Signal (MRS) is a pilot signal that is used for subframe synchronization and identification of downlink beams. It is transmitted every 5\,ms and cycles through all 48 beams over a period of 20\,ms. Although there are significant differences such as the subcarrier bandwidth, it can be used similarly to the LTE Reference Signal Received Power (RSRP).
For each period, the highest of the 48 MRSRP values is used as $P_R$. The cable loss between the UE's RF frontend and the UE's antenna can be considered as part of the constant $G_R$, which then becomes 4\,dBi. On the transmitter side, we can directly use the configured power $P_T$ of the AAS, because the radio chains and power amplifiers are directly connected with the antenna elements. However, depending on where the UE is located with respect to the AAS, a directive antenna gain $G_T(\phi,\theta)$ is considered. Because this is a beamforming antenna with a grid of beams, an artificial envelope pattern can be calculated by cycling through all beams sequentially, then virtually overlay them, and taking the maximum at each angle, see \figurename~\ref{fig:antenna_patterns}. The total measurement uncertainty is 6.25\,dB.\looseness=-1

The UE has been moved around with a van, or where driving was not possible, pushed by hand. When using the van, the UE and scanner antennas were mounted 16\,cm over the vehicle's roof to limit its influence on the antenna characteristic, which resulted in 2.1\,m above ground. When manually pushing the UE around, the antennas were at a height of 1.4\,m above ground, corresponding to the approximate height at which an adult is usually holding a smartphone.

\subsection{Comparison With 4G/LTE}
For the analysis of the 5G cell, it is interesting to have a comparable 4G (LTE) cell available. Thus, the test deployments were chosen to have not only an existing macro site antenna mast for installation of the 5G AAS, but also to have a live LTE signal on a legacy frequency band available.

\begin{figure}
	\setlength\abovecaptionskip{-0.3\baselineskip}
	\centering
	\includegraphics[width=\linewidth]{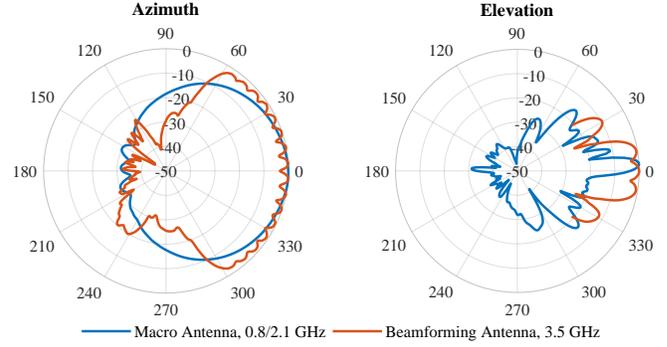}
	\caption{Antenna pattern envelope for a macro antenna and the AAS.\label{fig:antenna_patterns}}
	\vspace{-0.5em}
\end{figure}

For the measurements, a mobile network scanner (Rohde \& Schwarz\textsuperscript{\textregistered} TSMW) with a laptop running Nemo Outdoor has been used to log propagation channel metrics based on the LTE downlink signal. The RF was locked to one frequency band (800\,MHz or 2.1\,GHz) and all RSRP values per cell were recorded. For the analysis, only the values of the cell(s) of interest were used. The 2$\times$2~MIMO antennas (Rohde \& Schwarz\textsuperscript{\textregistered} TSMW-Z8) which were used, emulate omnidirectional smartphone antennas to give realistic results. Together with the cable losses, their gain is considered with -3\,dBi each (note that smartphone antennas are usually integrated and partially covered, leading to negative antenna gains). On the LTE base station side, the transmit power is corrected with the feeder cable losses in order to have the transmit power $P_T$ with respect to the antenna ports. Again, a directive antenna gain $G_T(\phi,\theta)$ is considered (see \figurename~\ref{fig:antenna_patterns}), depending on the relative location of the testbed UE to the LTE BS antenna. The total measurement uncertainty is 4.5\,dB.\looseness=-1

The network scanner has been fitted inside the 5G testbed UE, and the 2$\times$2~MIMO antennas could also be mounted close to the 5G antennas of the UE, without shielding them.

\subsection{Rural Area Environment}
The rural deployment was located close to a village (Meikirch) north-west of Bern, Switzerland.
The selected live macro site only had an LTE 800\,MHz cell with a 10\,MHz carrier directed over flat open fields towards a village at 1--1.5\,km distance.
The 5G AAS was installed right below the live macro antenna at a height of 12.4\,m above ground, with the same azimuth angle as the macro antenna. The vertical separation of the macro antenna to the testbed AAS was 2.1\,m, which resulted in a mean elevation angle difference of 0.22$^\circ$ seen from the UE antenna over all measurement sample locations.
The maximum used transmit power for 3.5\,GHz was 1560\,$\mathrm{W_{ERP}}$.

The measurements took place in mid-July 2018 during a sunny and clear week with an air temperature of around 22$^\circ$C.

\subsection{Suburban Area Environment}
\label{sec:suburban}
The suburban deployment was done in Ittigen, just north-east of Bern, Switzerland, situated on a slightly ascending slope.
For comparison, the signal from an existing LTE macro site at 2.1\,GHz with a 20\,MHz carrier was used.
The testbed AAS has been mounted on a smaller extra mast next to the LTE macro site mast due to space limitations on the latter. The height of the AAS was 24.5\,m above ground. The horizontal separation between the antennas was 3.8\,m, while the vertical separation was 8.5\,m. The mean angular separation of both antennas seen from the UE over all measurement sample locations is 0.232$^\circ$ in azimuth and 0.819$^\circ$ in elevation and is regarded as negligible.
The maximum transmit power for 3.5\,GHz had to be limited due to stringent non-ionizing radiation (NIR) regulations in Switzerland. Without lowering the transmit power on the live macro cells, the allowed maximum transmit power for 3.5\,GHz was 384\,$\mathrm{W_{ERP}}$.

The measurements took place end of April and beginning of May 2018, with sunny and also overcast days, but no rain. The air temperature was around 15--20$^\circ$C.

\subsection{Urban Area Environment}
For an urban deployment, a suitable and existing live macro site was selected in the city of Zurich, Switzerland.
The AAS has been mounted just below a macro sector antenna on the same mast at a height of 29.4\,m above ground, overlooking a flat part of the city. Unfortunately, due to technical issues, no measurements of an LTE legacy band could be taken for this scenario.
Also here, the NIR regulations required that we had to lower the maximum transmit power. By slightly reducing the power on a few live macro cells on this site, we were allowed to use the same maximum 384\,$\mathrm{W_{ERP}}$ for the 3.5\,GHz 5G cell as in the suburban deployment.

The measurements took place middle of August 2018 during a sunny week with an air temperature of around 25$^\circ$C.

\subsection{Outdoor-to-Indoor Scenario}
Additionally to outdoor drive and walk tests, indoor measurements were collected at four buildings of the suburban deployment and two buildings of the urban deployment. The goal was to assess the penetration loss into buildings for frequencies above 3\,GHz. Measurements were started inside, just behind a window with line-of-sight to the 5G AAS. The UE was then slowly and steadily moved further inside the building, either until the connection dropped or the back side of the building was reached. Outdoor reference measurements were either taken just in front of the building or on a terrace on top of the building.

\section{Measurement Results}
\label{sec:measresults}

For all environments described in the previous section, a total of 267 data sets were analyzed. Each site deployment was analyzed separately, and the path loss was obtained according to (\ref{eq:pl_meas_based}).
Using a least squares (LS) regression, the path loss exponent $\gamma$ is estimated with the reference distance $d_0$ at 100\,m, and the standard deviation $\chi_\sigma$ is calculated according to the model (\ref{eq:simpl_pl_model}). The estimated values are listed in Table~\ref{table:pl_model_params}.
A comparison with the models SUI IEEE~802.16 (A, B, C), ECC-33, WINNER~II (C1, C2, D1), and 3GPP (RMa, UMa) was performed. In general, the ECC-33 model overestimates the path loss and is not well suited because it was derived by fitting curves for distances of 1 to 10\,km. For the few models that agree the most with the respective measurements, prediction error statistics have been computed. The mean prediction error $\mu_e$ indicates an over- (positive value) or under-prediction (negative value) with standard deviation $\sigma_e$, and the root mean square of the prediction error RMSE represents the general metric for comparison of how well the models fit the measurements. These were calculated in log-domain [dB].\looseness=-2

\begin{table}
	\centering
	\caption{Estimated Path Loss Model Parameters for (\ref{eq:simpl_pl_model})}
	\label{table:pl_model_params}
	\vspace{-0.8em}
	\begin{tabular}{@{}l|cc|cc|cc@{}}
		\toprule
Frequency & \multicolumn{2}{c}{Rural LOS/NLOS} & \multicolumn{2}{c}{Suburban} & \multicolumn{2}{c}{Urban} \\
 & $\gamma$ & $\chi_\sigma$ [dB] & $\gamma$ & $\chi_\sigma$ [dB] & $\gamma$ & $\chi_\sigma$ [dB] \\
\midrule
3.55\,GHz & 2.3 / 3.1 & 5.1 / 9.4 & 2.9 & 6.9 & 4.8 & 7.1 \\
2.1\,GHz  & --  & --  & 3.7 & 7.0 & --  & -- \\
800\,MHz  & 2.8 / 3.4 & 5.9 / 7.5 & -- & -- & --  & -- \\
		\bottomrule
	\end{tabular}
	\vspace{-0.5em}
\end{table}

\subsection{Rural Environment}

\begin{figure}[b]
	\setlength\abovecaptionskip{-0.2\baselineskip}
	\centering
	\includegraphics[width=\linewidth]{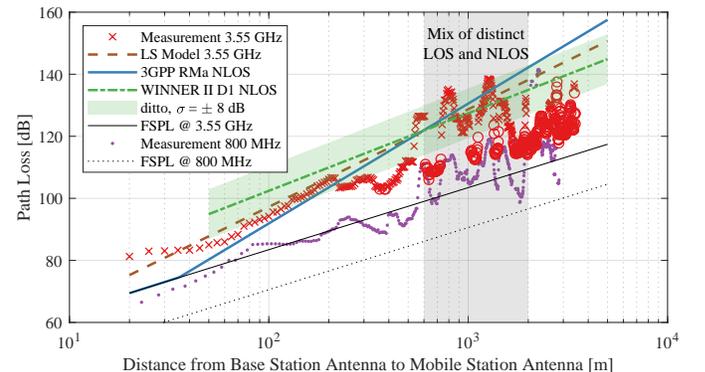}
	\caption{Rural environment measurements compared to models.\label{fig:PL_rural}}
\end{figure}

As expected, this environment provides a good share of line-of-sight (LOS) opportunities, in fact 42\,\% of the measurement samples are estimated to be in LOS condition. The large share of LOS makes it necessary to analyze LOS and NLOS separately.
The classification for LOS/NLOS has been done by manually defining five coordinate polygons. All measurement samples within these five polygons were categorized as LOS, the others as NLOS.
The measured path loss for 3.5\,GHz is shown in \figurename~\ref{fig:PL_rural} (LOS marked with circles) with predictions from the models 3GPP Rural Macro (RMa) NLOS and the WINNER~II D1 (rural macro) NLOS. The free space path loss (FSPL) is also shown as a reference.
Variations in the path loss from 600\,m to 2\,km (see shaded area) show the changes between LOS and NLOS environments.
All error statistics are listed in Table~\ref{table:pl_error_stats}. Both models overestimate the path loss by 14.9\,dB and 9.8\,dB, respectively, and the WINNER~II D1 rural NLOS model fits best, although the measurements are often well outside of its 8\,dB standard deviation.
The shadow fading distribution is shown in \figurename~\ref{fig:PLdist_all} in the top two distributions for LOS and NLOS.
\looseness=-1

\begin{figure}
	\setlength\abovecaptionskip{-0.2\baselineskip}
	\centering
	\includegraphics[width=\linewidth]{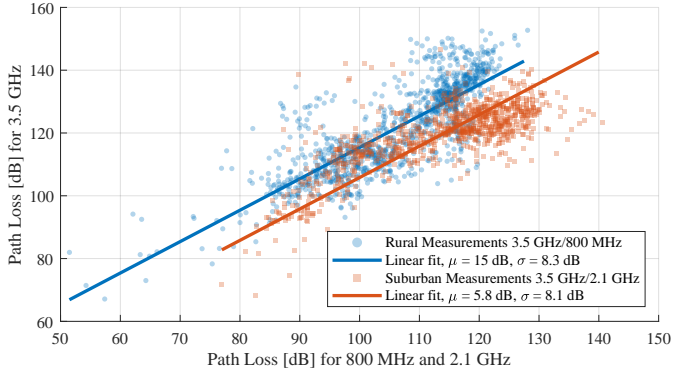}
	\caption{Path loss for 3.5 GHz vs. 800 MHz and 2.1 GHz at each measurement location; validates the theory of $12.8\,\mathrm{dB}$ and $4.4\,\mathrm{dB}$ path loss difference.\label{fig:PL35vsLTE}}
	\vspace{-1em}
\end{figure}

For comparison, the measured path loss for the 800\,MHz LTE signal is also shown in \figurename~\ref{fig:PL_rural}, with the corresponding FSPL as a reference. Furthermore, for each 5\,m measurement grid location, both path losses are plotted in \figurename~\ref{fig:PL35vsLTE}. Assuming only a frequency dependent offset, a line with a slope of $10\,\mathrm{dB}/10\,\mathrm{dB}$ can be fitted. This offset in theory is $20\,\log_{10}(3.5\,\mathrm{GHz}/0.8\,\mathrm{GHz})=12.8\,\mathrm{dB}$, but measurements show 15\,dB with a standard deviation of 8.3\,dB. The 2.2\,dB residual is within the measurement uncertainty.\looseness=-1

\begin{figure}
	\setlength\abovecaptionskip{-0.2\baselineskip}
	\centering
	\includegraphics[width=\linewidth]{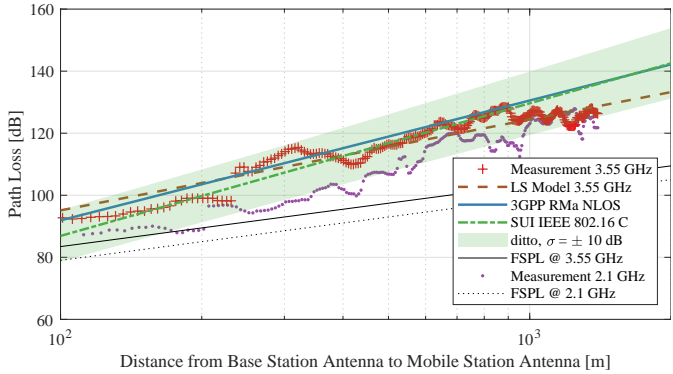}
	\caption{Suburban environment measurements compared to models.\label{fig:PL_suburban}}
\end{figure}

\subsection{Suburban Environment}

As explained in Section~\ref{sec:suburban}, parallel measurements at 2.1\,GHz have been conducted for this campaign.
During validation of the data, two anomalies were corrected: the first stemmed from a loss of connectivity because of a large apartment complex affecting signal propagation. The affected street segment was excluded from the analysis; the second stemmed from the local geography, resulting in a much higher than expected signal strength. Fitting the data to a two-ray model \cite{goldsmith_wireless_2005} using the delay spread information from the network scanner allowed us to explain this anomaly. The corrected overall path loss versus distance is shown in \figurename~\ref{fig:PL_suburban}.
The environment does not offer many LOS opportunities. We could therefore categorize LOS areas by manually defining coordinate polygons. As a result, 14\,\% of all samples fall in such a polygon.

\begin{figure}
	\setlength\abovecaptionskip{-0.2\baselineskip}
	\centering
	\includegraphics[width=\linewidth]{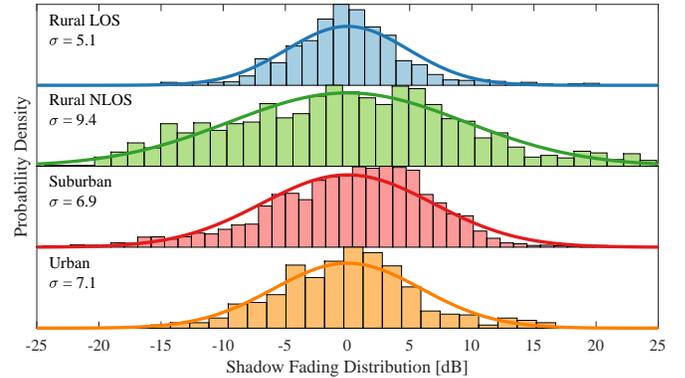}
	\caption{Distribution of the measured shadow fading components. The solid curve represents the Gaussian fit for each distribution and is an indication if the shadow fading has a log-normal distribution.\label{fig:PLdist_all}}
\end{figure}

Comparing empirical models with the measurements at 3.5\,GHz shows the best agreement with the SUI Terrain C model (RMSE~=~4.87\,dB), and the 3GPP RMa NLOS (RMSE~=~5.29\,dB) model. Both models overestimate the path loss by 2.11\,dB and 3.83\,dB, respectively. These results are also listed in Table~\ref{table:pl_error_stats}.
The shadow fading with a close to normal distribution is shown in \figurename~\ref{fig:PLdist_all} on the third row.

The available measurement data on 2.1\,GHz also allows us to do comparisons. The measured path loss at 2.1\,GHz is plotted in \figurename~\ref{fig:PL_suburban} and a comparison with 3.5\,GHz is shown in \figurename~\ref{fig:PL35vsLTE}. Fitting a slope of $10\,\mathrm{dB}/10\,\mathrm{dB}$ results in an offset of 5.8\,dB with a standard deviation of 8.1\,dB. In theory, this frequency dependent offset is $20\,\log_{10}(3.5\,\mathrm{GHz}/2.1\,\mathrm{GHz})=4.4\,\mathrm{dB}$. The 1.4\,dB residual is within the measurement uncertainty.

\begin{figure}
	\setlength\abovecaptionskip{-0.2\baselineskip}
	\centering
	\includegraphics[width=\linewidth]{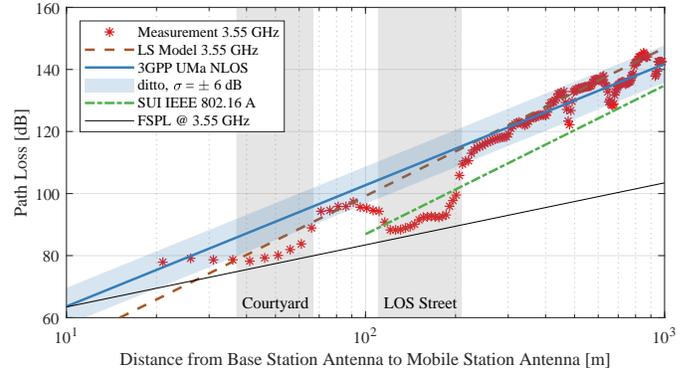}
	\caption{Urban environment measurements compared to models.\label{fig:PL_urban}}
	\vspace{-0.5em}
\end{figure}

\subsection{Urban Environment}

In our measurement data from the urban deployment, we have 27\,\% of the samples in LOS conditions. This may seem a lot, but considering a dense deployment with inter-site distances of a few hundred meters and only considering outdoor coverage, this becomes realistic. The main LOS contributions come from one long street at the azimuth angle of the antenna, between 110-210\,m (see the gray-shaded area in \figurename~\ref{fig:PL_urban}). Another area with close to LOS path loss has been identified as the courtyard of the city block on which the antenna was mounted (south-west corner, overlooking the city block towards north-east).
Regarding the comparison with empirical models, the two that fit the best, are the 3GPP Urban Macro (UMa) NLOS (RMSE~=~6.84\,dB) and the SUI Terrain A (RMSE~=~9.78\,dB). The 3GPP UMa model slightly overestimates the path loss by 2.1\,dB, and the SUI Terrain A model underestimates the path loss by -8.3\,dB. Also these results are listed in Table~\ref{table:pl_error_stats}.
The shadow fading distribution is shown at the bottom of \figurename~\ref{fig:PLdist_all} and conforms to the normal distribution.\looseness=-1

\subsection{Outdoor-to-Indoor Scenario}

\begin{figure}
	\setlength\abovecaptionskip{-0.2\baselineskip}
	\centering
	\includegraphics[width=\linewidth]{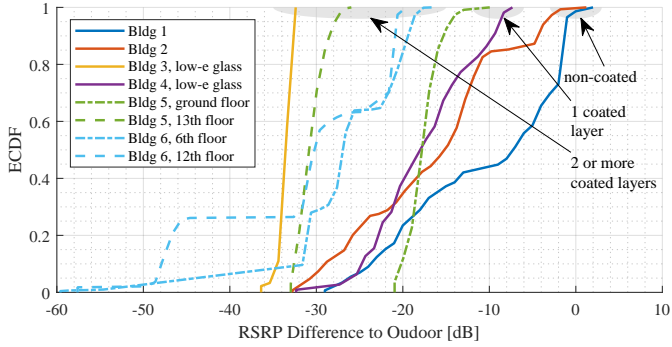}
	\caption{Outdoor-to-indoor penetration loss shown as empirical CDF of the received power difference from indoor compared to outdoor.\label{fig:RSRPindoor-outdoor}}
	\vspace{-1em}
\end{figure}

The indoor RSRP measurements are normalized with the corresponding reference outdoor RSRP and the empirical cumulative distribution function (CDF) is plotted, see \figurename~\ref{fig:RSRPindoor-outdoor}. The buildings 1--4 are from the suburban deployment, the buildings 5 \& 6 from the urban deployment.
The measurements clearly show which buildings are new or with coated low-e windows, and which ones use regular windows:
building~1 is an older office building with regular double glazing windows.
Building~2 is a home for elderly people, also with uncoated double glazing windows.
Building~3 is a new apartment building, with triple glazing low-e windows, causing around 30\,dB of additional loss.
Building~4 is a newer office building with triple glazing windows and one coated layer.
Building~5 is a renovated 14 floor office building. The measurements from the ground floor behind large single layer shop-windows show the lower attenuation than from the 13th floor behind multiple glazing low-e windows.
Building~6 is a 12 floor renovated office building with low-e windows. The measurements from the 6th and 12th floor show about the same attenuation, only walls inside the building cause different variations in the path loss.
In general, we can say that newer or renovated buildings with coated low-e windows add 10--30\,dB of additional penetration loss compared with buildings with regular windows.\looseness=-1

\begin{table}
	\centering
	\caption{Path Loss Model Prediction Error Statistics}
	\label{table:pl_error_stats}
	\vspace{-0.8em}
	\begin{tabular}{@{}l|cc|cc|cc@{}}
		\toprule
		Environment & \multicolumn{2}{c}{Rural} & \multicolumn{2}{c}{Suburban} & \multicolumn{2}{c}{Urban} \\
		Model & RMa & WINNER2 & RMa & SUI C & UMa & SUI A \\
		\midrule
		$\mu_e$ [dB] & 14.9 & 9.82 & 3.83 & 2.11 & 2.07 & -8.3 \\
		$\sigma_e$ [dB] & 10.8 & 7.53 & 3.66 & 4.4 & 6.54 & 5.18 \\
		RMSE & 18.4 & 12.4 & 5.29 & 4.87 & 6.84 & 9.78 \\
		\bottomrule
	\end{tabular}
\end{table}

\section{Conclusions}
\label{sec:conclusions}

Measurements from extensive rural, suburban, and urban measurement campaigns with a 5G testbed operating in the 3.5\,GHz band have been analyzed and compared with predictions from empirical path loss models (3GPP, WINNER~II, SUI). Almost all models tend to overestimate the path loss. Even though no model gives the least error in all environments, the 3GPP group of models always ranked among the top two, with an over prediction of 2.1\,dB and 3.8\,dB for the urban and suburban scenario, and 14.9\,dB for the rural scenario.
Furthermore, the excessive attenuation of around 10\,dB per coating layer of low-e windows has an impact on outdoor-to-indoor penetration as the number of newly built and renovated houses grows.
Finally, path loss exponents are slightly lower with a beamforming antenna at 3.5\,GHz, compared to a conventional macro antenna at 2.1\,GHz and 800\,MHz.

\bibliographystyle{IEEEtran}
\bibliography{references}

\end{document}